\documentclass[twocolumn,aps,prc,showpacs,groupedaddress]{revtex4}
\usepackage{graphicx}
\usepackage{dcolumn}
\usepackage{bm}
\newcommand{\eqdef}[1]{\label{eq:#1}}
\newcommand{\eqref}[1]{Eq.(\ref{eq:#1})}

\begin{document}

\title{Solving the Richardson equations for Fermions}

\author{S.~Rombouts}
\author{D.~Van Neck}
\affiliation{Universiteit Gent, Vakgroep Subatomaire en Stralingsfysica,
             Proeftuinstraat 86, B-9000 Gent, Belgium}
\author{J.~Dukelsky}
\affiliation{ Instituto de Estructura de la Materia, CSIC, Serrano 123, 28006 Madrid, Spain}
\date{\today}

\begin{abstract}
Forty years ago Richardson showed that the eigenstates of the pairing Hamiltonian with constant
interaction strength can be calculated by solving a set of non-linear coupled equations. However, in
the case of Fermions these equations lead to singularities which made them very hard to solve. This
letter explains how these singularities can be avoided through a change of variables making the
Fermionic pairing problem numerically solvable for arbitrary single particle energies and
degeneracies.
\end{abstract}

\maketitle

\section*{Introduction}

Exactly solvable models serve as a guideline for understanding the properties of correlated
many-body systems. Already in the sixties, Richardson solved the eigenproblem of a constant pairing
interaction in a set non-degenerate single-particle levels for Fermion~\cite{Rich63} and
Boson~\cite{Richa} systems. However, it turned out that in the Fermion case the solutions exhibit
singularities which are hard to treat numerically~\cite{Rich64,Rich65}. Recently, exactly solvable
pairing models have gained new attention~\cite{Heri01}, with applications to nanometalic
grains~\cite{Sierra} (for a review see \cite{Delf01}), Bose Einstein Condensates~\cite{Duke01c}
and nuclear physics~\cite{Duke01b}. 
It was first shown that the pairing model was integrable 
by finding the complete set of commuting integrals of motions \cite{Camb97}, 
and subsequently, three new families of fully integrable and 
exactly solvable models \cite{Duke01a} giving rise to a large class 
of pairing Hamiltonians with non-uniform matrix elements \cite{Dukec} were presented. 
These models are exactly solvable, except for the singularities occurring in Fermion systems 
for some critical values of the pairing strength. 
This problem, in spite of some early attempts to cure it \cite{Rich66a,Ha}, 
precluded for over forty years the use of these exactly solvable models 
for a wide range of applications, ranging from
condensed matter to nuclear physics. Moreover, the recent developed extensions of the exact solution
to pairing Hamiltonians including the isospin degree of freedom ($O(5)$ pairing) \cite{Pan} with
promising applications to $N \sim Z$ nuclei and high $T_c$ superconductivity, suffer from the same
kind of singularities.

In this letter show how the Richardson equations can be solved numerically, avoiding the
singularities, through an appropriate change of variables. This procedure provides a fast an
accurate way to solve the equations for a constant pairing interaction. The method can be
applied as well to more general exactly solvable Hamiltonians 
associated with a coupled set of non-linear equations of the Richardson type.

\begin{figure}
\begin{center}
\includegraphics[width=8.6cm]{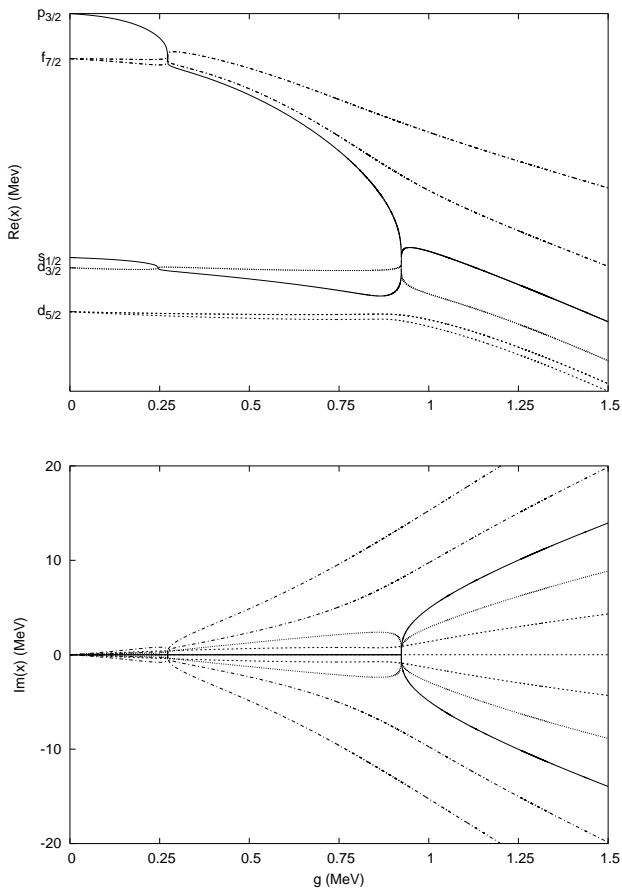}
\caption{Real and imaginary part of the variables $x_\alpha$ for the model described in the text
\label{fig:xvars}}
\end{center}
\end{figure}

The exactly solvable pairing Hamiltonian has the following form: 
\begin{equation}
 H =   \sum_{j,m} e_j a_{jm}^{\dagger}a_{jm} 
    \\  - \frac{g}{4} \sum_{j,m,j',m'} a_{jm}^{\dagger} a_{j\bar{m}}^{\dagger} a_{j'\bar{m}'} a_{j'm'},
\eqdef{pairinghamiltonian}
\end{equation}
with $e_j$ any set of single-particle energies and $g$ the pairing interaction strength.
The $N$-pair eigenstates of this Hamiltonian have the form
\begin{equation}
 \prod_{\alpha=1}^{N} \left[ \sum_{j,m} \frac{1}{2e_j - x_\alpha}  
                      a_{jm}^{\dagger} a_{j\bar{m}}^{\dagger} \right] |0\rangle,
\eqdef{bethe_ansatz}
\end{equation}
where $|0\rangle$ is a state without paired particles (a Racah quasispin vacuum state, \cite{Ring80}).
The corresponding energy $E(\{x\})$ is given by
\begin{equation}
E(\{x\})= \langle 0 | H | 0 \rangle + \sum_{\alpha} x_\alpha.
\end{equation}
The complex variables $x_\alpha$ can be found by solving a set of non-linear equations:
\begin{equation}
 \sum_j \frac{d_j}{2e_j - x_\alpha} + \sum_{\beta=1, \beta \neq \alpha}^{N} 
                      \frac{1}{x_\beta - x_\alpha} + \frac{1}{2g} = 0,
 \eqdef{richardson}
\end{equation}
for $\alpha=1,\ldots,N$. The parameters $d_j$ depend on the level degeneracies and on the structure
of the vacuum state $|0\rangle$. For Fermions, $ d_j=\frac{\nu_j -\Omega_j}{2}, $ where $\Omega_j$
is the pair degeneracy (for the nuclear shell model $ \Omega_j = j + 1/2 $ ), and $\nu_j$ is the
seniority of the level $j$. Note that $d_j \leq 0$ because of the Pauli principle. Solving this set
of nonlinear equations solves the eigenproblem for the pairing
Hamiltonian~\eqref{pairinghamiltonian}. Unfortunately, the algebraic solution becomes numerically
unstable at certain critical values of the interaction strength~\cite{Rich64,Rich65}. This is caused
by singularities in the first and second terms in \eqref{richardson}, when some of the variables
$x_{\alpha}$ are approaching the value $2e_j$. One can understand this from the electromagnetic
analogy for the exactly solvable pairing model~\cite{Duke02}: in the Fermion case, the
single-particle levels $e_j$ and the variables $x_\alpha$ correspond to opposite charges. Therefore
a group of variables can cluster around a single-particle level in such a way that for each of the
variables the repulsive charge of the other variables is compensated for by the attractive charge of
the single-particle level. These singularities occur for Fermions in double or multiple degenerate
levels. In the case of doubly degenerate equidistant levels these singularities can be handled for
the ground state~\cite{Rich66a}, but the problem arises again in the treatment of the excited states
\cite{Roman}. In previous calculations it was necessary to tune the solutions by hand as soon as
they approach a singularity. No general solution method was known until now.
Figure~\ref{fig:xvars} illustrates the behavior of the variables in case of multiply degenerate levels
(see below for the details of the model).

A general approach to solve the Richardson equations \eqref{richardson},
starts from an approximate solution in the weak-interaction limit (see below).
Then this solution is evolved adiabatically up to the desired interaction strength 
by gradually increasing the value of $g$.
At each step in $g$, the previous solution has to be updated.
For dealing with the singularities, it is useful to notice first 
that the variables $x_\alpha$ in \eqref{richardson}\ 
are most sensitive to the other variables nearby.
Therefore one can divide the set of variables $x_\alpha$ into 
several {\em clusters} of variables, 
grouped around different single-particle levels.
By solving the equations for each cluster separately,
one can obtain a solution for the whole system iteratively.
%
%
A practical way to organize the clusters,
is to link each variable to its nearest single-particle level $2 e_j$,
and to consider a cluster for each level that has variables around it.

The question now is how to solve the equations for each cluster,
particularly in the case of singularities.
Let us consider the set of indices $C_k$ of the $N_k$ variables that cluster around a level $2 e_k$.
Then one can consider the equations
\begin{equation}
 \frac{d_k}{2 e_k - x_\alpha}  + \sum_{\beta \in C_k, \beta \neq \alpha} \frac{1}{x_\beta - x_\alpha}
  + F_k(x_\alpha) = 0, \ \ \ \ \forall \alpha \in C_k,
 \eqdef{clusterequation}
\end{equation}
with
\begin{equation}
  F_k(x)= \frac{1}{2g} + \sum_{j,j \neq k}  \frac{d_j}{2 e_j - x} +
           \sum_{\beta, \beta \notin C_k } \frac{1}{x_\beta - x}.
 \eqdef{clusterfk}
\end{equation}
The function $F_k(x)$ describes the influence of the other levels and the variables of the other
clusters on the variables in the cluster $C_k$. Because of the way the clusters are set up, the
function $F_k(x)$ will be a smooth function in the region around $e_k$ where the variables of the
cluster are located. The singularities will occur in the first two terms of \eqref{clusterequation}.
In the case that some of the variables in the cluster approach the value $2 e_k$, the divergences in
the first and the second term of \eqref{clusterequation}\ must cancel out. Multiplying
\eqref{clusterequation}\ by $2 e_k - x_\alpha$, and summing over the $n_k$ variables $x_\alpha$ at
the singular point, leads to the condition
\begin{equation} 
  n_k = - 2 d_k +1,
\eqdef{singularity_condition}
\end{equation}
with $n_k$ the number of variables that actually converge to $2 e_k$.
For Fermions the value $(-2 d_k)$ corresponds to the pair degeneracy of the level $e_k$:
because of the Pauli principle, no more Fermion pairs can occupy the level.
Trying to put more pairs in that level results in a singularity.
In fact, the structure of the ground state \eqref{bethe_ansatz}\ does not
result in a forbidden occupation of the level.
However, on expanding the wave function of \eqref{bethe_ansatz}\
in terms of the pair creation operators  $a_{jm}^{\dagger} a_{j\bar{m}}^{\dagger}$,
one finds that the leading term cancels out because of the Fermionic anti-commutation rules.
This translates into the numerical difficulties encountered in the solution of the equations.
For Bosons $d_k>0$, and hence \eqref{singularity_condition}\ shows
that singularities do not occur in the Bosonic case.
This can also be understood from the electrostatic analogy: 
Boson pairs and single-particle levels have charges of the same sign.
Therefore the variables try to avoid each other at all times,
and singularities do not occur.

The above procedure suggests a way to remove the singularities from the equations:
multiplying  \eqref{clusterequation}\ with $(2 e_k - x_\alpha)^p$, for some power $p$,
and summing over all variables $x_\alpha$ in the cluster.
The resulting equations become,
%
for $p>1$:
\begin{equation}
 \left(d_k + N_k - \frac{p}{2}\right) S_{p-1}  
   + \frac{1}{2} \sum_{k=2}^{p-1} S_{k-1} S_{p-k}
  + R_p = 0,
 \eqdef{clusterequation3}
\end{equation}
with
\begin{eqnarray}
 S_p  & = & \sum_{\alpha \in C_k}  (2 e_k - x_\alpha)^{p} 
 \\
 R_p  & = & \sum_{\alpha \in C_k}  (2e_k - x_\alpha)^{p}  F_k(x_\alpha).
\end{eqnarray}
The compact form of \eqref{clusterequation3}\ suggests that it might be
advantageous to solve them for the new variables $S_p$ instead
of the original variables $x_\alpha$.
Note that given a set of variables $S_1,\ldots, S_{N_k}$, 
one can easily construct the polynomial whose roots correspond
to the values $2e_k - x_\alpha$.
Hence one can switch from one set of variables to the other.
The problem comes with the quantities $R_p$:
these are functions of the $x_\alpha$, and it is not straightforward
to express them as functions of the $S_p$.
However, for a given set of variables $x_\alpha$, 
one can easily evaluate the values $S_p$ and $R_p$.
Furthermore, one can evaluate the gradient matrix $G$,
with $G_{lm}$ the derivative with respect to $S_m$ 
of \eqref{clusterequation3}\ for $p=l+1$,
\begin{equation}
 G = G^S+G^R,
\end{equation}
where
\begin{eqnarray}
 G^S_{lm} & = & \left\{ \begin{array}{ll} 
                    d_k + N_k - \frac{l+1}{2}, & \ \ \mbox{for $m=l$} \\
                    S_{l-m},                   & \ \ \mbox{for $m<l$} \\
                      0,                       & \ \ \mbox{for $m>l$} 
			\end{array} \right. \\
  G^R_{lm} & = & \frac{d R_{l+1}}{d S_m} 
              = \sum_{\alpha \in C_k} \frac{d R_{l+1}}{d x_\alpha} \frac{d x_\alpha}{d S_m},
\end{eqnarray}
for $l,m=1,\ldots,N_k$.
$G^R$ can be evaluated accurately using a special 
inversion algorithm for Vandermonde matrices~\cite{Golu89}.
Therefore one can solve the new set of equations,
\eqref{clusterequation3}\ for $p=2,\ldots N_k+1$, 
in the new variables $S_1, \ldots S_{N_k}$
using a standard gradient technique such as the 
multi-dimensional Newton-Raphson method~\cite{Pres92}. 
However, in the case of a singularity, the gradient matrix
becomes ill-conditioned: 
the diagonal elements of the gradient matrix are given by 
\begin{equation}
 G^S_{ll}= d_k + N_k - \frac{l+1}{2},
\end{equation}
for $l=1, \ldots N_k$.
The diagonal element will vanish for the index $ l_s = 2(d_k + N_k) -1$.
This will occur as soon as $N_k \geq -2 d_k +1$, 
which matches the value for which singularities can occur, see \eqref{singularity_condition}.
In such a case the lower-triangular matrix $G^S$ becomes singular.
The other part of the gradient matrix, $G^R$,
is derived from the smooth function $F_k$.
A series expansion of $F_k(x)$ in $x$ will be dominated by the lowest orders.
Therefore the elements of $G^R_{lm}$ are very small for larger values of $m$.
As a result, the value of $S_{l_s}$ can not be determined accurately from the set
of equations \eqref{clusterequation3}.
One can avoid this problem by limiting the cluster sizes to at most
the critical value $N_k= n_k = -2 d_k +1$,
and by using $g S_{-1}$ as an unknown variable instead of $S_{N_k}$.
If more variables are found near to the same single-particle level,
one can always divide the cluster into smaller, well-separated clusters, 
because at most $n_k$ of the variables can approach the single-particle level closely.
Knowing $S_{-1}$ and $S_1, \ldots, S_{N_k-1}$, one can still
straightforwardly construct the polynomial whose roots give
the corresponding values $x_\alpha$. 
Therefore one can easily switch between the two sets of variables.
Furthermore $g S_{-1}$ behaves smoothly, even at a singularity.
To set up an efficient gradient method,
it is useful to replace \eqref{clusterequation3}\ for the last value,
$p=N_k+1$, by a similar equation obtained using $p=0$:
\begin{equation}
 d_k g S_{-1} + g R_0 = 0,
 \eqdef{clusterequation3b}
\end{equation}
with $R_0=\sum_{\alpha \in C_k} F_k(x_\alpha)$.

For weak interaction strengths the function $F_k(x_\alpha)$ 
is dominated by the constant term $\frac{1}{2g}$, see \eqref{clusterfk}.
In the weak interaction limit one can take $F_k$ to be a constant.
The resulting functions $R_p$ take the simple form
\begin{equation}
 R_p= \frac{S_p}{2g}.
\end{equation}
Now the equations \eqref{clusterequation3}\  and \eqref{clusterequation3b}\
can be solved straightforwardly to yield the variables $S_l$,
from which one can construct the polynomial that gives a unique set
of variables $x_\alpha$.
The resulting eigenstate will depend on the size of the cluster
for each of the single-particle levels.
This establishes a one-to-one correlation between the eigenstates
of the non-interacting system ($g=0$) and the eigenstates of the
weakly-interacting system. 
One can conclude that the Richardson equations are complete:
their solutions generate all eigenstates,
and there are no spurious solutions.

One can obtain the solutions for strong interaction strengths
by solving the weakly interacting case first, 
and then gradually increasing the interaction strength.
At each step one can use the gradient method outlined above
in order to update the solution to the new interaction strength.
It is useful to adapt the stepsize in interaction strength 
to the convergence of the iterative procedure by taking smaller
steps in $g$ when the convergence of the Newton-Raphson method
for the variables $S_i$ becomes slower, which typically occurs
around the critical $g$ values.
One more ingredient is needed to avoid problems with the singularities:
when the interaction strength passes through a critical value,
the variables $x_\alpha$ passing through a singularity
can change from real to complex or vice versa.
At the same time the variables $S_l$ will become very small,
except for $l=-1$.
Even using the new variables, 
the gradient method does not lead to the right solution 
when it has to pass through critical values of the interaction strength.
One can avoid this problem by including an extrapolation step
based upon the previous solutions.
This extrapolation has to be done in the variables $S_l$, 
because they vary smoothly through the singularities and they remain real all the time.
Assume that converged solutions $x_\alpha'$ and $x_\alpha''$ were 
obtained for values $g'$ and $g''$ of the interaction strength.
The variables $x_\alpha''$ are grouped into clusters.
For each cluster, one evaluates the variables $S_p'$ and $S_p''$. 
Because the variables $S_p$ behave smoothly, even near a singularity,
one can estimate the variables $S_p$ for the new interaction strength $g$
by linear extrapolation:
\begin{equation}
 S_p = \frac{(g-g') S_p'' - (g-g'') S_p' }{g''-g'}
\end{equation}
The resulting values of $S_p$ can then be updated 
using the Newton-Raphson method or another gradient method.
The extrapolation step avoids the problems with the singularities
and greatly improves the convergence of the method.

\begin{figure}
\begin{center}
\includegraphics[width=8.6cm]{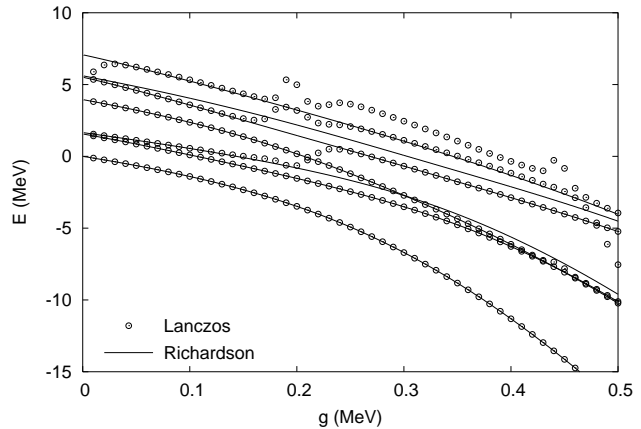}
\caption{Lanczos and Richardson results for the energies of the lowest zero-seniority states,
relative to the non-interacting ground-state energy.
\label{fig:glevels}}
\end{center}
\end{figure}

\begin{table}[h]
\centering
\begin{ruledtabular}
  \begin{tabular} {|c|cccc|}
   \hline
        level  &   energy        &  degeneracy &  pair occupation& seniority  \\
               & $ e_j (MeV)$    & $\Omega_j $ &  (ground state) & $ \nu_j $  \\
   \hline
   \hline
   $1d_{5/2}$   &     -21.5607   &      3      &      3     &     0      \\
   $1d_{3/2}$   &     -19.6359   &      2      &      2     &     0      \\
   $2s_{1/2}$   &     -19.1840   &      1      &      1     &     0      \\
   \hline
   $1f_{7/2}$   &     -10.4576   &      4      &      4     &     0      \\
   $2p_{3/2}$   & \    -8.4804   &      2      &      1     &     0      \\
   $1f_{5/2}$   & \    -7.7003   &      3      &      0     &     0      \\
   $2p_{1/2}$   & \    -7.6512   &      1      &      0     &     0      \\
   \hline
   $3s_{1/2}$   & \    -0.3861   &      1      &      0     &     0      \\
   $2d_{5/2}$   & \ \   0.2225   &      3      &      0     &     0      \\
   $1g_{9/2}$   & \ \   0.5631   &      5      &      0     &     0      \\
   \hline
  \end{tabular}
 \end{ruledtabular}
  \caption{Woods-Saxon single-particle levels~\cite{Romb98}.}
  \label{table:splevels}
\end{table}
As an example, consider the level scheme listed in Table~\ref{table:splevels},
together with a constant pairing interaction.
This model describes neutrons in $^{56}$Fe; 
its ground state and finite-temperature properties 
have been studied using a Quantum Monte Carlo method~\cite{Romb98}.
The eigenstates can also be found through the solution
of Richardson's equations, \eqref{richardson}, or through Lanczos
diagonalization in a seniority basis~\cite{Voly01}.
The full many-body space has a dimension of the order of $10^{15}$,
while the zero-seniority basis has dimension 14894.
In Fig.~\ref{fig:glevels} the lowest zero-seniority eigenvalues for this model are
shown as a function of the interaction strength,
calculated using the Lanczos method and using the method explained above 
(only 50 Lanczos iterations were used).
\begin{figure}
\begin{center}
\includegraphics[width=8.6cm]{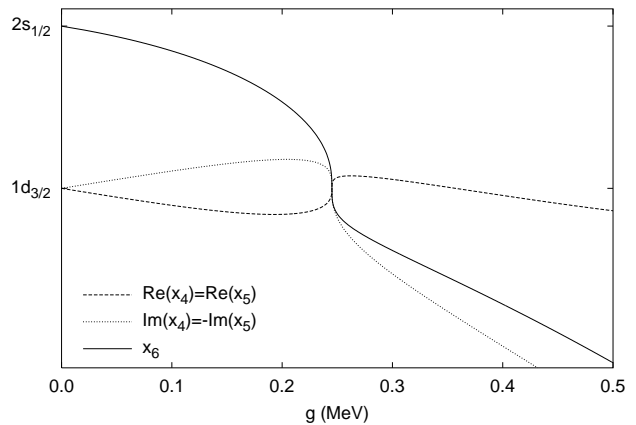}
\caption{Behavior of the variables near the singularity around the $2 d_{3/2}$ level. 
Variables $x_4$ and $x_5$ are complex conjugates, $x_6$ is real over the whole range
of $g$ values.
\label{fig:xsingular}}
\end{center}
\end{figure}
\begin{figure}
\begin{center}
\includegraphics[width=8.6cm]{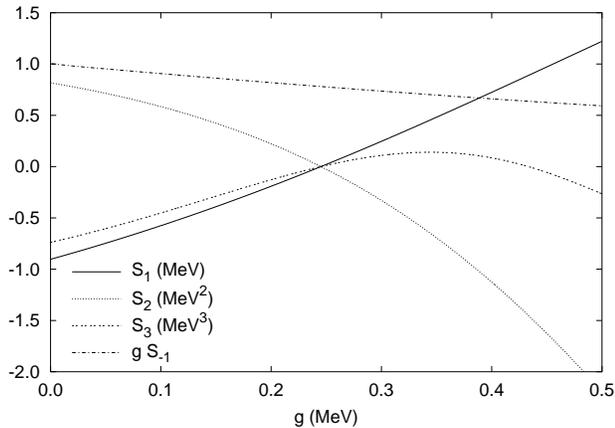}
\caption{Smooth behavior of the new variables near the $2 d_{3/2}$ singularity.
\label{fig:ssingular}}
\end{center}
\end{figure}
To calculate the ground state,
a straightforward implementation of the Newton-Raphson method for the original
equations, \eqref{richardson}, works well up to an interaction strength of $g \simeq 0.2$.
It turns out that a singularity occurs around the $1d_{3/2}$ level at a value of $g=0.245$.
A Fortran-95 computer program was written based upon the procedure outlined above.
It was able to solve the equations for all interaction strengths in a matter of seconds.
Figure~\ref{fig:xsingular} shows the behavior of the three $x$ variables
that cluster around the $1d_{3/2}$ level as a function of $g$.
In Fig.~\ref{fig:ssingular} one can see that the corresponding variables $S_l$
behave much more smoothly.
More singularities occur around other levels at higher interaction strengths,
as is shown in Fig.~\ref{fig:xvars}.
The solution of the Richardson equations is faster than the Lanczos method,
it requires less computer memory, and it works for all the eigenstates,
not just the ground state. 
Moreover, the procedure can deal with much larger systems, well beyond the limits
of large-scale exact diagonalizations.

This work shows that the exactly solvable pairing models are indeed solvable
in practice, even for Fermions with multiple degeneracies. 
It opens up a whole new range of applications for these models. 
Furthermore, similar techniques might be useful to solve the non-linear
equations for other exactly solvable pairing and spin models~\cite{Duke01a,Feng98,Pan,Shas}

\acknowledgments

We wish to thank R.~Richardson, S.~Pittel, J.~Draayer, F.~Pan and K.~Heyde 
for the interesting discussions and suggestions. 
This work was supported by the Fund for Scientific Research - Flanders (Belgium), 
the Research Board of Ghent University and by the Spanish DGI Grant BFM2003-05316-C02-02.


\end{document}